\documentclass[prl,a4paper,twocolumn,showpacs,superscriptaddress,amsmath,amsfonts,amssymb,floatfix,preprintnumbers]{revtex4}
\pdfoutput=1
\usepackage[T1]{fontenc}\usepackage[latin1]{inputenc}
\usepackage{dcolumn,graphicx,color,booktabs} \graphicspath{{Images/}}
\usepackage[charter]{mathdesign}
\usepackage{amsmath,bm}
\renewcommand{\figurename}{Fig.}
\makeatletter\renewcommand{\fnum@figure}[1]{\figurename~\thefigure~(color online).}\makeatother
\definecolor{DarkBlue}{rgb}{0,0,0.5}

\begin{document} \pagestyle{plain}

\title{Bridging charge-orbital ordering and Fermi surface instabilities \\in half-doped single-layered manganite
La$_{0.5}$Sr$_{1.5}$MnO$_4$}

\author{D.\,V.~Evtushinsky}\author{D.\,S.~Inosov}
\affiliation{Institute for Solid State Research, IFW Dresden, P.\,O.\,Box 270116, D-01171 Dresden, Germany}
\author{G.~Urbanik}
\affiliation{Institute for Solid State Research, IFW Dresden, P.\,O.\,Box 270116, D-01171 Dresden, Germany}
\affiliation{Institute of Experimental Physics, University of Wroclaw, pl. Maxa Borna 9, 50-204 Wroclaw, Poland}
\author{V.~B.~Zabolotnyy}\author{R.~Schuster}\author{P.~Sass}\author{T.~Hänke}
\affiliation{Institute for Solid State Research, IFW Dresden, P.\,O.\,Box 270116, D-01171 Dresden, Germany}
\author{C.~Hess}
\affiliation{Institute for Solid State Research, IFW Dresden,
P.\,O.\,Box 270116, D-01171 Dresden, Germany}
\author{B.~Büchner}
\affiliation{Institute for Solid State Research, IFW Dresden,
P.\,O.\,Box 270116, D-01171 Dresden, Germany}

\author{R.~Follath}
\affiliation{BESSY GmbH, Albert-Einstein-Strasse 15, 12489 Berlin, Germany}
\author{P.~Reutler}\author{A.~Revcolevschi}
\affiliation{Laboratoire de Physico-Chimie de l'Etat Solide, Universit\'{e} Paris-Sud XI, 91405 Orsay C\'{e}dex, France}
\author{A.\,A.~Kordyuk}
\affiliation{Institute for Solid State Research, IFW Dresden, P.\,O.\,Box 270116, D-01171 Dresden, Germany}
\affiliation{Institute of Metal Physics of National Academy of Sciences of Ukraine, 03142 Kyiv, Ukraine}
\author{S.\,V.~Borisenko}
\affiliation{Institute for Solid State Research, IFW Dresden, P.\,O.\,Box 270116, D-01171 Dresden, Germany}

\begin{abstract}

\end{abstract}


\maketitle
\noindent


\textbf{Density waves are inherent to the phase diagrams of
materials that exhibit unusual, and sometimes extraordinarily useful
properties, such as superconductivity and colossal magnetoresistance
\cite{Littlewood1, Tokura_Nagaosa, Morosan, Davis_checkerboard}.
While the pure charge density waves (CDW) are well described by an
itinerant approach \cite{nesting, Sergey_TaSe}, where electrons are
treated as waves propagating through the crystal, the charge-orbital
ordering (COO) is usually explained by a local approach
\cite{Khomskii1}, where the electrons are treated as localized on
the atomic sites. Here we show that in the half-doped manganite
La$_{0.5}$Sr$_{1.5}$MnO$_4$ (LSMO) the electronic susceptibility,
calculated from the angle-resolved photoemission spectra (ARPES),
exhibits a prominent nesting-driven peak at one quarter of the
Brillouin zone diagonal, that is equal to the reciprocal lattice
vector of the charge-orbital pattern. Our results demonstrate that
the Fermi surface geometry determines the propensity of the system to
form a COO state which, in turn, implies the applicability of the
itinerant approach also to the COO.}

\begin{figure}[h!]
\vspace{-0ex}\includegraphics[width=1\columnwidth]{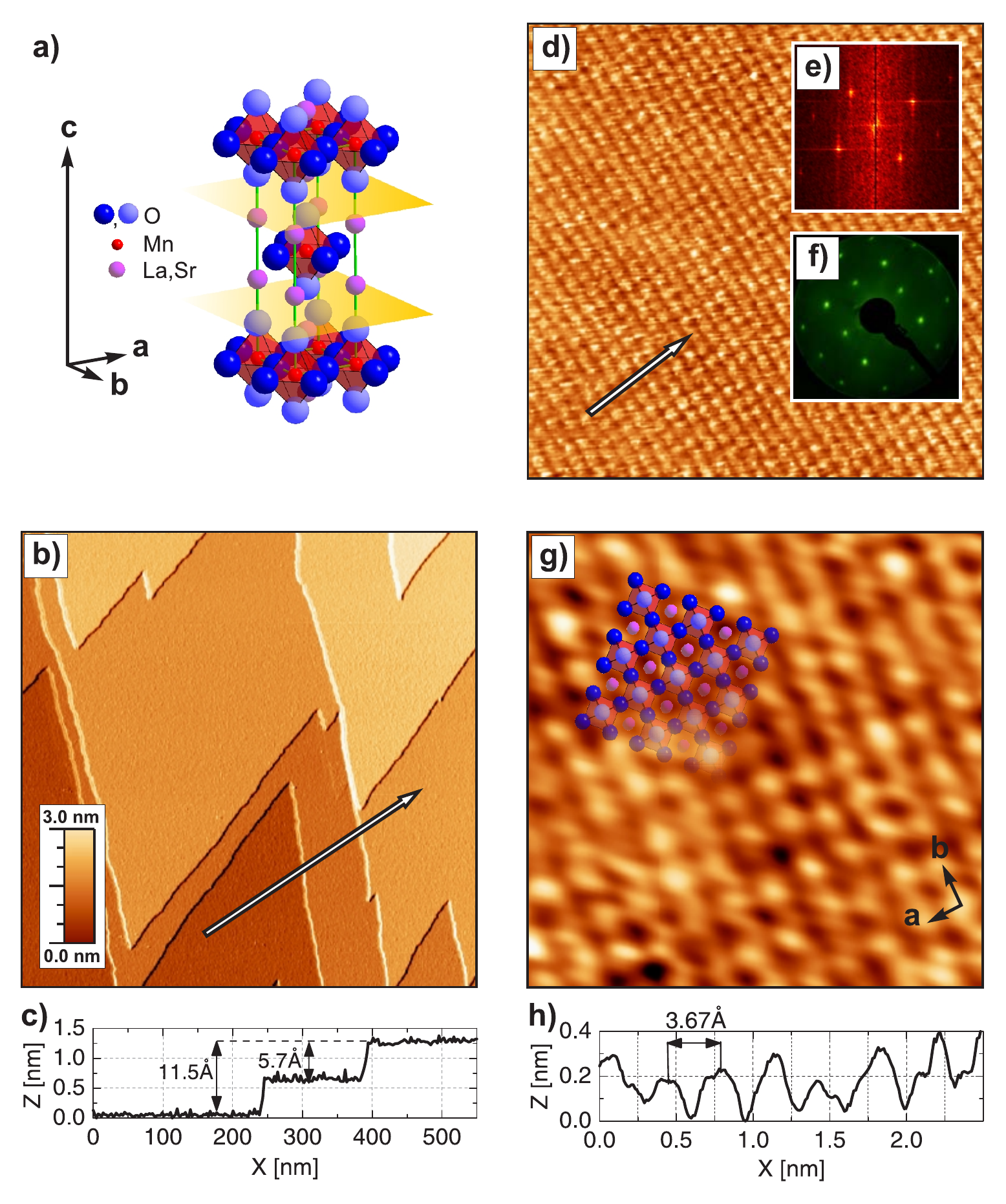}\vspace{-3ex} \caption{Crystalline structure and surface quality of La$_{0.5}$Sr$_{1.5}$MnO$_4$ single crystals. STM topographic images, taken at 300\,K. The sample is positively biased relative to the STM tip (V=0.5\,V, I=0.2\,nA).(a) The crystalline structure of La$_{0.5}$Sr$_{1.5}$MnO$_4$. The elementary cell is marked by green lines, and cleavage planes are shown in yellow. (b) $1\text{\,}\mu \text{m}\times1\text{\,}\mu \text{m}$ image of atomically flat micrometer-sized terraces separated by steps. (c) The line profile measured on the figure (b), highlighting half-unit-cell high steps. (d) $10\text{\,nm}\times10\text{\,nm}$ atomic-resolution image measured on the flat terrace from figure (b). (e) The Fourier transform of the STM image (d). (f) LEED image, taken at 100\,eV with the same orientation as the STM image, indicates a reconstruction-free surface. (g) Higher-magnification image of an area from the figure (d). (h) The line profile measured on the figure (d), showing the atomic modulations. The distance between the adjacent maxima corresponds to the value of the lattice constant.\vspace{-1em} }
 \label{f:Model1}
\end{figure}

\begin{figure}[t]
\includegraphics[width=\columnwidth ]{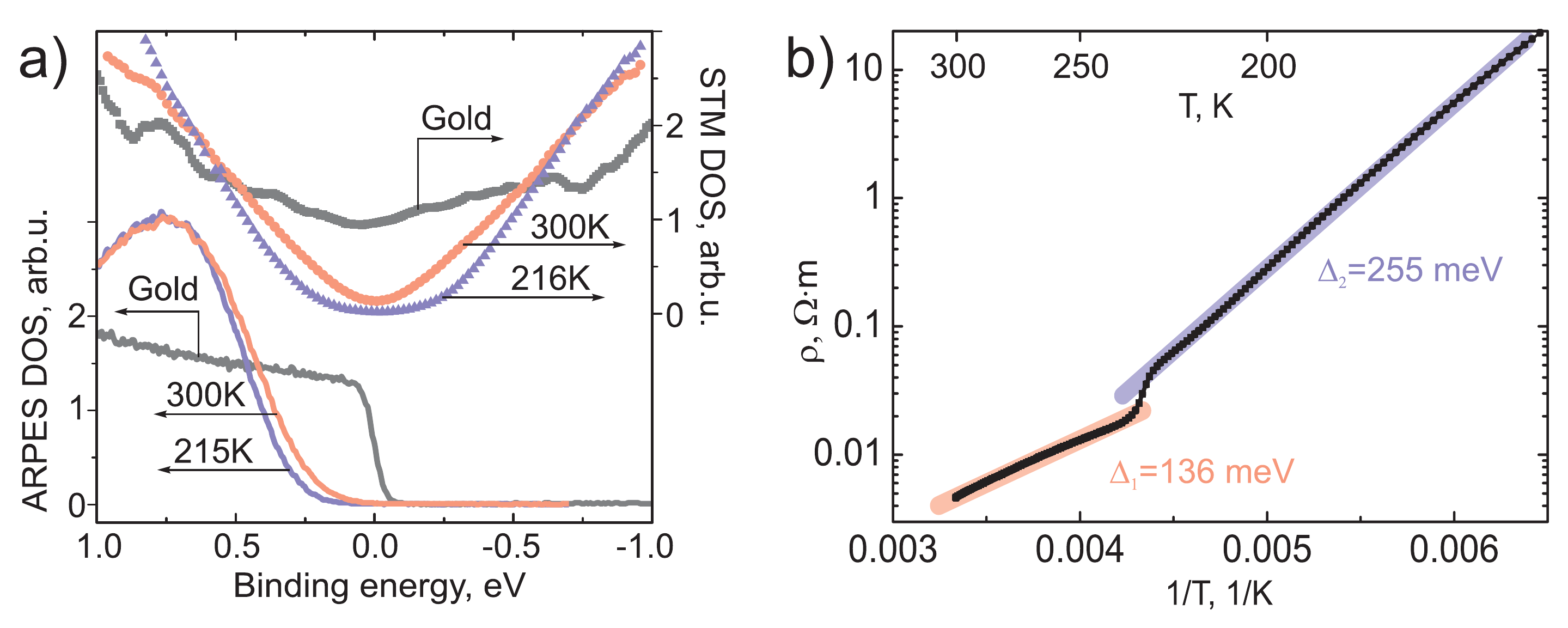}
\caption{The similarity of the surface and bulk electronic
structures is indicated by the consistency among the conductive
properties and behavior of the DOS near the
Fermi level. (a) The DOS extracted from the tunneling (averaged over a
whole momentum space) and photoemission ($\Gamma$-pocket) data above
and below $T_{\text{COO}}$. Spectroscopies on gold are shown for
calibration. (b) $\ln\rho$ plotted versus $1/T$. Linear fit above
and below $T_\text{COO}$ is shown as underlying thick lines.}
\vspace{-0ex}
\label{f:Model2}
\end{figure}

\begin{figure*}[]
\includegraphics[width=\textwidth ]{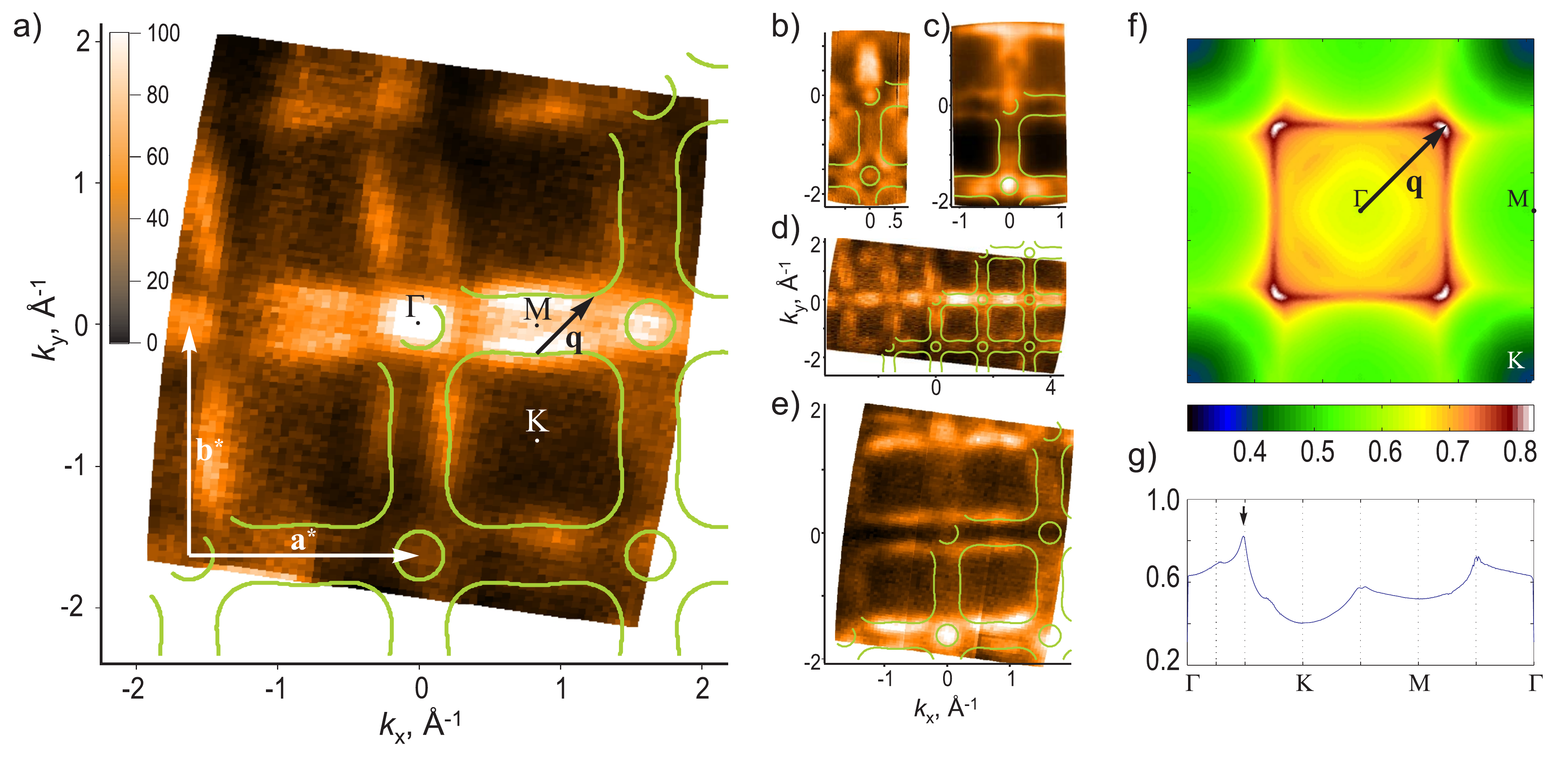}
\caption{Electronic structure and susceptibility of La$_{0.5}$Sr$_{1.5}$MnO$_4$. Band dispersion is derived from ARPES spectra, taken at different experimental conditions, to neutralize effect of matrix elements. The electronic susceptibility, calculated on the basis of the revealed band dispersion, shows a prominent peak at the vector $\mathbf{q}=\mathbf{a^*}/4+\mathbf{b^*}/4$. The photoemission intensity, integrated in energy window of 70\,meV width, centered at 190\,meV below Fermi level, is shown in panels (a)\,--\,(e) for different excitation energies and light polarizations: (a) linear horizontal, 157\,eV, 300\,K; (b) linear horizontal, 37\,eV, 190\,K; (c) linear horizontal, 70\,eV, 195\,K; (d) linear horizontal, 200\,eV, 300\,K; (e) linear vertical, 157\,eV, 300\,K. Contours of the remnant Fermi surface, which consists of a small round electron-like pocket centered at the $\Gamma$ point and a large square hole-like barrel centered at the K point, are presented in panels (a)\,--\,(e) by green lines. The calculated  susceptibility is shown in panel (f), and its profile along the high-symmetry directions is shown in (g).}
\vspace{-0ex}
\label{f:Model3}
\end{figure*}

\begin{figure}[]
\includegraphics[width=\columnwidth ]{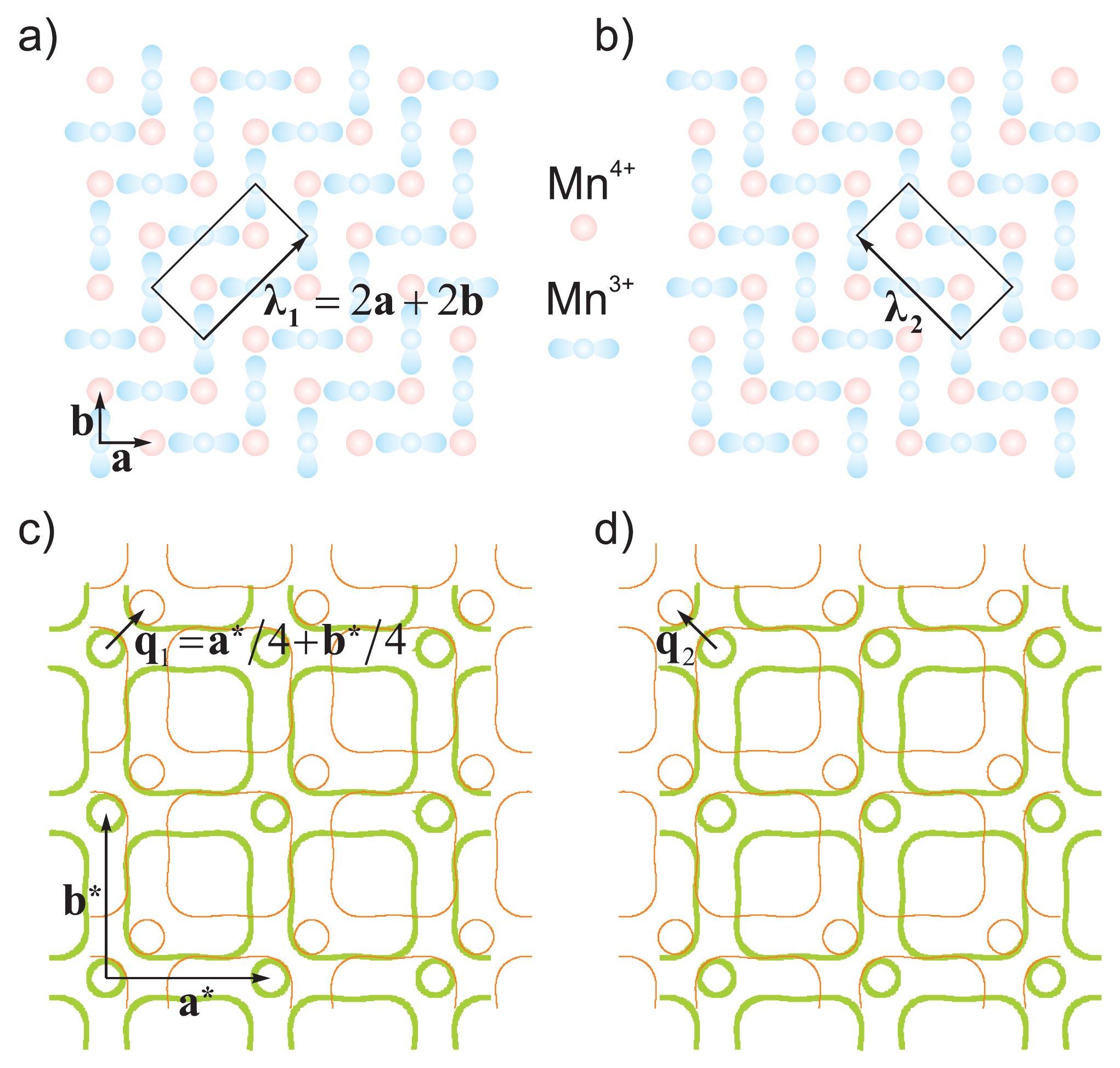}
\caption{The modulation in direct space and the corresponding nesting in reciprocal space; illustration of two possible orientations of the charge-orbital pattern. (a), (b) Real-space pattern of the charge-orbital order with lattice vectors $\bm{\lambda_{1,2}}=\pm2\mathbf{a}+2\mathbf{b}$ along zigzags. (c), (d) Fermi surface (green solid lines) and its replicas (thin orange lines) displaced by the vectors $\mathbf{q_{1,2}}=\pm\mathbf{a^*}/4+\mathbf{b^*}/4$. Note that the observed nesting vector, $\mathbf{q_{1,2}}$ corresponds to the periodicity of the charge-orbital pattern along zigzags, $\bm{\lambda_{1,2}}$: $|\mathbf{q_{1,2}}|=(2\sqrt{2})^{-1}|\mathbf{a^*}|$, $|\bm{\lambda_{1,2}}|=2\sqrt{2}|\mathbf{a}|$, and $\mathbf{q_{1,2}} \parallel \bm{\lambda_{1,2}}$.}
 \label{f:Model4}
\end{figure}

In the present study we apply three different techniques, angle-resolved photoemission spectroscopy (ARPES), scanning tunneling microscopy (STM), and transport measurements, to exactly the same single crystals of
La$_{0.5}$Sr$_{1.5}$MnO$_4$ (LSMO), a renown representative of manganites, which exhibits a prominent transition to the charge-orbital ordered (COO) state \cite{magnetic_resistivity, nature_surface, neutrons, Reutler}. The STM results, summarized in Fig.~1, reveal high quality of the cleaved surface. Fig.~1(b) shows a typical STM topographic image
obtained on an \textit{in situ} cleaved surface of the manganite.
The measured step heights are equal to $n\times(0.62\pm0.02)$\,nm,
$n\in\mathbb{Z}$, that is a multiple of half the \textbf{c} lattice
constant [see Fig.~1(b), (c)]. This implies that the cleavage takes
place between the adjacent La,Sr oxide layers [see Fig.~1(a)]. Note
that in contrast to Ref.~\onlinecite{G1}, our data reveal well
resolved surfaces and steps, both above and below the COO temperature,
$T_\text{COO}=230$\,K. The measured roughness of the terraces,
observed in our data, was usually 0.1\,nm, indicating previous
assessments (0.6\,nm) \cite{nature_surface, G1} to be largely
overestimated. A large defect-free portion of the surface is revealed by the atomically resolved STM image is shown
in Fig.~1(d). The atomically resolved topography is highlighted by the Fourier transform of the STM image [Fig.~1(e)]. The absence of additional peaks in low-energy electron diffraction (LEED) image [Fig.~1(f)] confirms the absence of any significant surface reconstruction. Fig.~1(g) shows a higher-magnified image with a superimposed cartoon of the crystalline structure. To the best of our knowledge, this is the first real-space
observation of the La,Sr oxide layer in single-layered manganites.

In Fig.~2(a) we show both, photoemission and tunneling spectroscopic
data taken from the surface discussed above. The extracted densities
of states (DOS), shown on the same energy scale, exhibit similar
behaviour near the Fermi level. This semiconducting behaviour is in
agreement with the temperature dependence of the in-plane
resistivity $\rho$, which is linear in $\ln\rho$ versus $1/T$
coordinates [see Fig.~2(b)]. Remarkably, the values of the energy
gap, determined from the resistivity measurements
($\Delta_1=136$\,meV above and $\Delta_2=255$\,meV below
$T_\text{COO}$ respectively), are consistent with our spectroscopic
data, indicating that in LSMO the surface is a good representative
of the bulk. This is in agreement with the recent comparative X-ray
scattering studies of bulk and surface of the same material
\cite{nature_surface}.

Now we turn to the reciprocal space image of the LSMO electronic
structure. In Fig.~3(a)--(e) we present ARPES intensity maps. Since the spectral weight on the Fermi level is nearly absent [Fig.~2(a)], these maps represent the momentum distribution of the intensity
integrated in a finite energy window below the Fermi level (see caption to Fig.~3). In the following 
we will refer to these maps as remnant Fermi surface (FS) \cite{Shen_remnant}.
Since the energy gap, estimated from both spectroscopic and
transport data, is rather small in comparison with the bandwidth, the remnant FS is very close 
to the hypothetical FS of the non-gapped parent metal \cite{hypothetical_FS} and
thus has similar properties. As for each particular excitation energy the photoemission matrix
elements highlight and suppress different parts of the spectrum \cite{Sergey_matrix_elements, Dima_waterfalls}, which may lead to false conclusions as for the geometry of the electronic structure, we carried out the
series of measurements at different photon energies and light
polarizations, also analyzing data taken in different Brillouin
zones [see Fig.~3(a)--(e)]. The compilation of the data allows us to
conclude that the remnant FS of LSMO consists of a large hole-like barrel
centered at the K point and a small electron-like pocked centered at
the $\Gamma$ point [see Fig.~3(a)--(e)], just as in bilayer
manganites in their metallic state \cite{Dessau, Shen}. We found
that the remnant FS sheets are well described by the following tight binding
formula:
\begin{multline} \epsilon(k_x, k_y)=d_0+d_1(\cos k_xa+\cos k_ya)+d_2\cos k_xa\cos k_ya\\+d_3(\cos 2k_xa+\cos 2k_ya)
+d_4(\cos 2k_xa\cos k_ya \\+ \cos k_xa\cos 2k_ya)+d_5\cos 2k_xa\cos
2k_ya \end{multline} with $d_0=3.050$, $d_1=-4.638$, $d_2 = 3.022$,
$d_3=1.006$, $d_4=-1.720$, $d_5=0.660$ for the K-barrel, and
$d_0=5.33$, $d_1=-2.974$, $d_2=0$, $d_3=0$, $d_4=0$, $d_5=0$ for the
$\Gamma$-pocket, where $a=3.865$\,\AA\,\cite{Reutler} is a lattice
constant \cite{only_FS, no_T_dep}. It is interesting, that the area
enclosed by the contours (49\,\% of the Brillouin zone) is close to
the half-filling and since this will not change much in the absence
of the energy gap \cite{hypothetical_FS}, this is in contrast to the
nominal doping level of 1/2 electron per Mn atom, which corresponds to 25\,\% of the Brillouin zone. Here we also note, that at particular polarizations and excitation energies an apparent
anisotropy of the photoemission signal shows up [see, e.g. Fig.~3.(c),
(e)], which might point to the sample anisotropy in the a-b plane
\cite{anisotropy_Tokura}. However, rotation of the sample by
$90^{\circ}$ did not affect the measured signal, so we may conclude
that the mentioned anisotropy of the spectra arises purely from the
anisotropy of the photoemission matrix elements. The absence of any
inherent sample anisotropy both above and below $T_\text{COO}$ is in
agreement with the presence of small domains with different orientations
of the COO pattern \cite{domains}, see Fig.~4(a) and (b).

With the LSMO electronic structure at hand we try to understand the
origin of the charge-orbital ordering in this compound in the
framework of an itinerant approach. Thereto we refer to the experience
gained from studying compounds that exhibit another type of
ordering\,---\,charge density wave (CDW). In these compounds, the
major reason that drives the formation of the ordering is well
understood: if the electronic susceptibility (Lindhard function) of
the \textit{unreconstructed} system possesses a strong peak at a particular wave vector, the
system is likely to develop a density wave order at this vector \cite{nesting, Sergey_TaSe}. On the basis of the band dispersion extracted from ARPES data we have calculated the susceptibility of the hypothetical LSMO metal by a procedure similar to that described in Ref.\,\onlinecite{Dima_susceptibility}. The result of these calculations
is shown in Fig.~3 (f),\,(g). The susceptibility turnes out to have a
prominent nesting-driven peak very close to the vector
\begin{equation}
\mathbf{q}=\frac{\mathbf{a^*}}{4}+\frac{\mathbf{b^*}}{4},
\end{equation}
where $\mathbf{a^*}=2\pi\mathbf{a}/|\mathbf{a}|^2$ and
$\mathbf{b^*}=2\pi\mathbf{b}/|\mathbf{b}|^2$ are the lattice vectors
in the reciprocal (momentum) space, while $\mathbf{a}$ and
$\mathbf{b}$ are lattice vectors in the direct (coordinate) space \cite{incommensurability}.
Generally speaking, the FS nesting occurs when two replicas with
opposite Fermi velocities meet each other. It means that the nesting
is strong when FS sheets of the same origin (both hole-like or both
electron-like) are externally tangent, or when the FS sheets of
different origin (one hole-like and another electron-like) are
internally tangent. In Fig.~4(c), FS (green solid line) and its
replica shifted by the vector $\mathbf{q}$ (orange thin line) are
shown [panel (d) corresponds to a different orientation of the
charge-orbital pattern]. We see that the FS of LSMO is nested with
the vector $\mathbf{q}$\,---\,there are two regions that contribute
mostly to the peak in the susceptibility: the electron-like $\Gamma$-pocket
fits inside the rounded corner of the square-shaped hole-like
barrel, and the convex corner of the hole-like barrel fits inside its
concave part.

The nesting vector we observe corresponds to the lattice vector of
the charge-orbital superstructure, directed along zigzag chains [see
Fig.~4(a)]
\begin{equation}
\bm{\lambda}=2\mathbf{a}+2\mathbf{b}.
\end{equation}
Let's prove that $\bm{\lambda}$ corresponds to the nesting vector
$\mathbf{q}$. Eqs.~(2) and (3) in conjunction with $\mathbf{a}
\parallel \mathbf{a^*}$ and $\mathbf{b} \parallel \mathbf{b^*}$
result in $\mathbf{q} \parallel \bm{\lambda}$. Taking the modulus from
Eqs.~(2) and (3), and keeping in mind that in our case
$|\mathbf{a}|=|\mathbf{b}|$ as well as
$|\mathbf{a^*}|=|\mathbf{b^*}|$, we get
$|\mathbf{q}|=(2\sqrt{2})^{-1}|\mathbf{a^*}|$ and
$|\bm{\lambda}|=2\sqrt{2}|\mathbf{a}|.$ Finally recalling that
$\mathbf{a^*}\cdot\mathbf{a}=2\pi$, yields $\mathbf{q}\cdot\bm{\lambda}=2\pi$.

The found correspondence hints that the propensity of the system to form a
COO state is determined by the FS instability due to the
nesting-driven peak in the electronic susceptibility. Furthermore, significant changes in the physical properties, observed in LSMO upon charge-orbital ordering \cite{magnetic_resistivity, neutrons}, are similar to those observed upon charge ordering  \cite{Naito, Sergey_TaSe, Morosan, PG_Hall} in classical CDW systems. Although CDW and COO are very similar, no relation between the COO and the fermiology has been established to date to
our knowledge, albeit some theoretical works have predicted a
possibility of the description of the COO phenomena via the band
approach \cite{band_approach} as an alternative to the existing
local approach \cite{Khomskii1}. Our observations provide such a
link by suggesting that FS instabilities due to either Peierls-like
or other mechanisms can serve as a possible origin of the COO.


The project is part of the FOR538 and was supported by the DFG under
Grants No. KN393/4 and BO1912/2-1. We thank R.Hübel for technical
support, A.~Narduzzo and R.\,Schneider for resistivity measurements.
ARPES experiments were performed using the ``$1^3$ ARPES'' end
station at the Berliner Elektronenspeicherring-Gesellschaft für
Synchrotron Strahlung m.b.H. (BESSY).


\begin{thebibliography}{00}
\bibitem{hypothetical_FS} Taking into account small magnitude of the gap ($\Delta\sim150$\,meV) and large values of the Fermi velocity ($v_\text{F}\sim4$\,eV$\cdot\text{\AA}$) in the case of LSMO, one may estimate the difference between the observed remnant FS and the hypothetical FS of the noninteracting metal as much smaller than $\Delta/v_\text{F}\sim0.04\text{\,\AA}^{-1}$.
\bibitem{only_FS} Values of the tight-binding coefficients were adjusted to yield the best approximation of the band dispersion in the vicinity of the Fermi level, i.e. the input required to calculate the response of the system.
\bibitem{no_T_dep} We observe no significant temperature dependence of the intensity distribution in the remnant FS maps.
\bibitem{incommensurability} Deviation of the revealed peak position from commensurate value $\mathbf{a^*}/4+\mathbf{b^*}/4$ is about 2\,\%, i.e. within error bars.
\end{thebibliography}
\end{document}